\documentclass[prc,preprint]{revtex4-1}
\usepackage{graphicx,color}
\usepackage{multirow}

\newcommand{\beq}{\begin{equation}}
\newcommand{\eeq}{\end{equation}}
\newcommand{\beqn}{\begin{eqnarray}}
\newcommand{\eeqn}{\end{eqnarray}}

\begin{document}
\title{Well funneled nuclear structure landscape: renormalization}
\author{A. Idini $^{1}$ } 
\email{andrea.idini@gmail.com}
\author{G. Potel$^2$} 
\email{gregory.potel@gmail.com}
\author{F. Barranco $^{3}$}
\email{barranco@us.es}
\author{E. Vigezzi $^{4}$}
\email{vigezzi@mi.infn.it}
\author{R.A. Broglia $^{5,6}$}
\email{broglia@mi.infn.it}%
\affiliation{$^1$ Department of Physics, University of Jyvaskyla, FI-40014 Jyvaskyla, Finland}
\affiliation{$^2$ Lawrence Livermore National Laboratory, USA}
\affiliation{$^3$ Departamento de F\'isica Aplicada III,
Escuela Superior de Ingenieros, Universidad de Sevilla, Camino de los Descubrimientos, 	Sevilla, Spain}		
\affiliation{$^4$ INFN Sezione di Milano, Via Celoria 16, I-20133 Milano, Italy }											
\affiliation{$^5$ Dipartimento di Fisica, Universit\`a di Milano,
Via Celoria 16, 
I-20133 Milano, Italy }
\affiliation{$^6$ The Niels Bohr Institute, University of Copenhagen, 
DK-2100 Copenhagen, Denmark }
\date{\today}

\begin{abstract}
A complete characterization of the structure of nuclei can be obtained by combining 
information arising from inelastic scattering, Coulomb excitation and $\gamma-$decay, together
with one- and two-particle transfer reactions. In this way it is possible to probe 
the single-particle and collective components of the nuclear many-body wavefunction
resulting from their mutual coupling and diagonalising the low-energy Hamiltonian. 
We address the question of how accurately such a description can account for experimental
observations.  It is  concluded that renormalizing empirically and on equal footing  bare single-particle 
and collective  motion in terms of self-energy (mass) and vertex corrections (screening), 
as well as  particle-hole and pairing  interactions through particle-vibration 
coupling allows theory to provide  an overall, quantitative account of the data. 
\end{abstract}

 \maketitle
\date{today}

Nuclear structure  is both a mature \cite{Bohr,Bohr_nobel,Mott_nobel} and a very  active field of research \cite{INPC,50years} , and time seems ripe to attempt a balance
of our  present, quantitative understanding of it.
Here we take up an aspect of this challenge  and try to answer to the question:  how accurately can 
theory  predict  structure observables in terms of single-particle and collective degrees of freedom and of their couplings?

In pursuing this quest one of two paths can be taken: 1) select one nuclear property, 
for example the single-particle spectrum, and study it throughout the mass table \cite{Tarpanov}; 2) select
a target nucleus $A$ which has been fully characterized through inelastic scattering and Coulomb
excitation ($A(\alpha,\alpha')A^*$), together with one-  ($A(d,p)A+1, A(p,d)A-1)$ and 
two- ($(A+2(p,t)A,A(p,t)A-2$)  particle transfer processes and study the associated,  complete nuclear 
structure information involving the island of  nuclei $A$, $A \pm 1$ and $A \pm 2$ 
in terms of the corresponding absolute differential
cross sections and decay transition probabilities.
Here we have chosen the second way and selected the group of nuclei
$^{118,119,120,121,122}$Sn involved in
the  characterization of the spherical, superfluid $^{120}$Sn target nucleus. 

Single-particle and collective vibrations constitute  the  basis states. The calculations are implemented  in terms of a SLy4 effective interaction and a $v_{14}(^1S_0) (\equiv v_p^{bare})$ Argonne pairing potential.
HFB provides an embodiment of the quasiparticle spectrum while  QRPA a realization of  density  ($J^{\pi} = 2^+,3^-,4^+$, $5^-$) and  spin 
($2^{\pm},3^{\pm},4^{\pm}$, $5^{\pm}$) modes. 
Taking into account  renormalisation processes (self-energy, vertex corrections, phonon renormalization and  phonon exchange) in terms of  the particle-vibration  coupling (PVC) mechanism,
the dressed particles  as well as the  induced pairing interaction $v^{ind}_p$ were calculated  (see \cite{EPJ};
see also \cite{Avd:1,Avd:2,litvinova,litvinova2,Ring,colo,mizuyama,gnezdilov}).
Adding $v^{ind}_p$ to the bare interaction $v_p^{bare}$, the total pairing interaction $v_p^{eff}$ was determined. With these 
elements, the Nambu-Gor'kov  (NG) equation was solved selfconsistently using Green's function techniques \cite{Schrieffer,Idini1,Idini2,Soma,vanneck:93}, and
 the parameters characterizing the renormalized quasiparticle states
 obtained. Within this framework, the 
quasiparticle energies $\tilde E_{\nu}$ are given by 
$\tilde E_{\nu} = \sqrt{(\tilde \epsilon_{\nu} - \epsilon_F)^2 + {\tilde \Delta}^2_{\nu}} \; .$
The  renormalised single-particle energy
$\tilde \epsilon_{\nu} - \epsilon_F = Z_{\nu} [(\epsilon_{\nu} - \epsilon_F) + \Sigma_{\nu}^{even}]$,
is written in terms of the HF  energy $\epsilon_{\nu}$ and of the
 even part of the normal self-energy, the quantity  $Z_{\nu}$ providing 
a measure of the single-particle character of the orbital $\nu$.
The state dependent pairing gap $\tilde \Delta_{\nu}  =  \tilde \Delta_{\nu}^{bare} + \tilde {\Delta}_{\nu}^{ind}$
 obeys the generalized gap equation 
\begin{equation}
\tilde \Delta_{\nu}  = - Z_{\nu}  \sum_{\nu'>0} \langle \nu'  \bar \nu' | v_p^{bare} + v_p^{ind} | \nu \bar \nu \rangle
N_{\nu'} \frac{\tilde \Delta_{\nu'}}{2\tilde E_{\nu'}},
\end{equation} 
where $N_{\nu} =\tilde  u_{\nu}^2 + \tilde  v_{\nu}^2$ and $Z_{\nu} = \left( 1 - \frac{\Sigma^{odd}_{\nu}}{\tilde E_{\nu}} 
\right)^{-1} = m/(m_{\omega})_{\nu}$, $\Sigma_{\nu}^{odd}$ being the odd part of the normal self-energy, while 
$m_{\omega}$ is the $\omega-$mass. The resulting values of $\tilde \Delta_{\nu}$
are shown in Fig.1. The contribution of $v_p^{bare}$ and $v^{ind}$ to $\tilde \Delta_{\nu}$ are about equal,
density modes leading to attractive contributions which are partially cancelled out by spin modes, as expected from 
general transformation properties  of the associated operators entering the particle-vibration coupling vertices \cite{Borti,baldo,viverit}. 
Theory  {(SLY4 +QRPA+ (PVC) REN+NG)} provides a quantitative  
account of the experimental value
($\Delta^{exp} \approx$ 1.45 MeV). 
It is to be noted that in carrying out  the above calculations  use has been made  of empirically renormalized  collective modes.
This is because SLy4 leads to little collective density vibrations (cf. Table 1, where, for concreteness, the bare 
QRPA results characterizing the 
low-lying $2^+$ of $^{120}$Sn are collected \cite{EPJ}, see also \cite{terasaki}), 
in keeping with the associated value of the effective mass  0.7 $m$.
In fact, collectivity is closely associated with a density of levels  ($\sim m^*$) consistent with an effective
mass $m^* = m_{\omega}m_k/m \approx m$. 
This  is achieved by coupling  the two-quasiparticle QRPA SLy4 solutions to 4qp doorway states made out of a 2qp uncorrelated
component  and an empirically tuned  QRPA collective mode  \cite{footnote_new}
    (see Fig. 2 of \cite{EPJ}, cf. also \cite{Bertsch}), an example of the fact that 
in a consistent  PVC renormalised description of  the nuclear structure, one has to treat (dress), on equal footing, all
degrees of freedom (i.e. single-particle and collective modes). In this way, 
not only  self-energy but also vertex corrections are consistently  included (sum rules conserving processes), and thus
the "bare" QRPA mode is properly clothed, bringing theory in overall
 agreement with experiment 
 (see Table 1,  second and third lines; for details cf. \cite{Barranco_tobe} as well as \cite{EPJ}). 

To test how robust the results displayed in Fig. 1 are, we have recalculated $\tilde \Delta_{\nu}$ as a function  of the three parameters 
{\bf \{ $m_k$, $v_p^{bare}$, $\beta_{J}(J^{\pi})$ \} }  \cite{Skyrme}\cite{footnote2}\cite{footnote_new}.
The results 
($|\Delta_{h11/2} - \Delta^{exp} |$) associated with the lowest  quasiparticle state $h_{11/2}$ are displayed in Figs. 2(a),(b) and (c). 
They provide evidence
of the fact  that 
a description based on the renormalisation of single-particle states and collective modes through PVC 
leads to a well funnelled nuclear structure landscape 
\cite{Frauen}, 
displaying a global minimum 
for values of the set of parameters  $\;$ {\bf\{ \}}  close to the empirical values:  effective mass $m_k \approx 0.7 m$,  bare pairing interaction strength $v_p^{bare}$ consistent
with $v_{14}(^1S_0) (G_0 \approx 0.22$ MeV, see \cite{footnote2})  and quadrupole deformation parameter    
$(\beta_2)_0 \approx (\beta_2)_{exp} \approx 0.13 $.


Similar conclusions  can be drawn from
the study of the dependence on $v_p^{bare}$ and $\beta_2$ of the quasiparticle  spectrum  associated with  the  
valence orbitals  ($h_{11/2},d_{3/2},s_{1/2},g_{7/2}$ and $d_{5/2}$), of the splitting of the multiplet of states $(h_{11/2} \otimes 2^+)_{15/2^-  - 7/2^-} $ 
and  of the $\gamma-$decay spectrum following Coulomb excitation, as can be seen  from Figs. 2(d)-(f), Figs.2(g)-(h) and Fig.3, respectively. 


Because of  the PVC mechanism, the different valence quasiparticle states undergo renormalization and 
fragmentation, phenomena which can be specifically probed with one-particle transfer reactions.
In Fig. 4(a) we display  the 
absolute differential cross sections associated with the reaction  $^{120}Sn(d,p)^{121}Sn(lj)$,
calculated  making use of the spectroscopic amplitudes associated with the  strongest populated fragments of the 
valence orbitals $h_{11/2}, d_{3/2},s_{1/2}$ and $d_{5/2}$ and of global optical parameters, 
in comparison with the experimental data \cite{Bechara}. 
 Theory provides  an overall,  quantitative , account of the experimental findings. To be noted that 
 the  agreement  found between the summed absolute differential cross sections associated with the almost
 degenerate state $3/2^+$ and $11/2^-$ (experimentally non resolvable \cite{Bechara}, while theoretically separated by 
 100 keV), results from a subtle incoherent combination of the 
 $ l=2, d \sigma_{1n}/d\Omega$  peak at $\theta_{CM} \approx 20^0$
 and of that of the $l=4$ one at $\theta_{CM}= 47^0$.

In discussing the $^{120}\text{Sn}(p,d)^{119}$Sn reaction we concentrate on the $d_{5/2}$ orbital, the most theoretically challenging 
of all of the valence single-particle strength functions.
This is because  this state, being further 
away from the Fermi energy ($\epsilon_{d_{5/2}} = -11.3$ MeV, $\epsilon_F \approx  -8 $MeV) than the 
other four valence orbitals (see Table 2), 
is   embedded in  a denser set of doorway states (of type $s_{1/2} \otimes 2^+, d_{3/2} \otimes 2^+,
g_{7/2} \otimes 2^+, h_{11/2} \otimes 3^-$, etc.), as compared to the other ones. Consequently, it can  undergo accidental degeneracy 
and thus conspicuous  fragmentation.  As seen from Table 3, although 
the calculated summed cross sections ($\sigma = 6.15 $ mb) agree, within experimental
errors, with observation (7.93 $\pm$ 2 mb), {theory} predicts an essentially uniform fragmentation of the   strength over an energy interval
of $\approx$ 760  keV, while  the data \cite{Dickey} is consistent with a concentration of the strength at an energy close to that 
of the lowest theoretical $5/2^+$ level (1090 keV).

In keeping with the above scenario we have shifted the  bare single-particle energy $\epsilon_{d_{5/2}}$ by 
600 keV ((-11.3 + 0.6)  MeV= - 10.7 MeV), amounting to a 6\% change in the $k-$mass (i.e. from 0.7$m$ to $0.74 m$, Table 2 ), and recalculated
all the quantities discussed above.
Making use of the corresponding  nuclear structure 
results and of global optical parameters, the absolute differential cross sections associated 
with the $5/2^+$ states populated in the reaction 
$^{120}Sn(p,d)^{119}Sn$ and lying below 2 MeV have been calculated. They are
displayed in Fig.  4(b) in comparison with the experimental data. Theory provides  now a quantitative account of the experimental findings. In particular, of  the fact that 
the strength function is dominated by a single peak. With the 600 keV shift, it is predicted at an energy of 1050 keV carrying 4.4 mb and it is observed at 1090 keV with 
a cross section of 5.35  $\pm 1.3 $ mb. The resulting overall 
 agreement between theory and experiment is further confirmed by  Fig. 4(c) where the absolute value of the one-particle transfer strength function
associated with  the population of $5/2^+$ states predicted by the calculation is compared with experiment. 
As observed in Fig. 4(d), also this two-dimensional projection of the multidimensional  nuclear structure landscape is  funnelled, testifying to the 
physical robustness of the findings. 

An  alternative approach to the one discussed above 
which leads to almost identical findings regarding the $d_{5/2}$ fragmentation, can be obtained by treating the energy of the five valence orbitals 
as parameters to be optimized selfconsistently within the framework  of the full NG calculations, so as to best reproduce the quasiparticle spectrum (Opt. Table 2).
The results can be expressed in terms of a state-dependent $k-$mass $\langle (m_k)_{\nu}/m \rangle \approx 0.74 \pm \sigma$, with $\sigma=0.07$.


Making again use of the effective occupation numbers resulting from the solution of the NG equation, the 
two-nucleon spectroscopic amplitudes of the reactions
$^{120}Sn(p,t)^{118}Sn(gs)$ and  $^{122}Sn(p,t)^{120}Sn(gs)$ have been calculated. 
With the help of these quantities  and of global optical parameters, the absolute differential 
cross sections  have  been calculated 
in second-order DWBA taking into account successive and simultaneous transfer, properly corrected from non-orthogonality contributions \cite{Potel_review}. 
They are displayed  in Figs  1(b) and 1(c)  in comparison with the experimental findings \cite{Guazzoni1,Guazzoni2}. 
Theory reproduces the absolute differential cross sections 
associated with the ground state transitions
 within experimental errors. The calculations have been repeated for different values of the strength of the PVC 
associated with the most important collective vibrational mode, namely  the lowest $2^+$ as well as for different strengths of the bare  pairing interaction.
While the dependence of $\sigma_{2n}(p,t)$ is very weak with $\beta_2$ (not shown),
it is conspicuous with $v_p^{bare}$.
 An example of such dependence  is displayed 
 in  the inset to Fig. 1(a). Again, this 
two-dimensional section of the  nuclear
structure landscape is of a well funnelled character. 
Within this context, it is noted that a measure of the reliability with which theory can describe  the nuclear structure  is provided by the relative dimensionless standard 
deviations $\sigma_{rel}$  (equal to e.g. $\sigma(E_{qp})/\langle E_{qp}\rangle $ in  the case  of the quasiparticle spectrum)
associated with each of the different observables, and taken at  the minimum of the nuclear structure landscape, as shown in Table 4.



 We conclude that a theoretical description of  nuclear structure based on single-particle (mean field
with $m_k \approx 0.7 m$) 
and collective motion (QRPA) and on their interweaving controlled 
by the  
particle-vibration coupling mechanism and leading to renormalization of both types of nuclear excitations through mass (self-energy) and screening (vertex) corrections
and induced pairing, can  provide an overall  quantitative account  
of the nuclear structure representative of a mass zone (group of nuclei displaying homogeneous properties like e.g. sphericity and superfluidity, 
likely circumscribed by phase transition domains). 
Allowing for a weak state dependence  of the $k$--mass, 
determined by optimising the energy of the valence single-particle orbitals to reproduce the quasiparticle spectrum,  the theoretical description of the   nuclear structure 
 probed in terms of  direct reaction absolute differential cross sections and based on renormalized single--particle and collective degrees of freedom, becomes accurate within a 10\% error level. The PVC mechanism  is found to  play a central role in achieving this result.
Within this context,  we note that pairing in typical superfluid nuclei lying along  the stability valley
like $^{118,119,120,121,122}$Sn has a dual origin, in which $v_p^{bare}$ and 
$v_p^{ind}$ contribute essentially equally to the pairing gap.
The above considerations and protocols  are not only   transferable to the remaining  superfluid Sn-isotopes (not considered explicitely in the present case), 
but also applicable  to the quantitative description of other spherical, superfluid nuclear mass zones. 
\section{Acknowledgments}
This work have been supported by the Academy of Finland and University of Jyv\"aaskyl\"a within the FIDIPRO program and by the Helmholtz Association through the Nuclear Astrophysics Virtual Institute (VH-VI-417) and the Helmholtz International Center for FAIR within the framework of the LOEWE program launched by the state of Hesse.

 \clearpage

 \begin{table}[h]
\begin{tabular}{|c|c|c|c|}
\cline{2-4}
\multicolumn{1}{c|}{}  & \quad $\hbar \omega_{2+} $ (MeV)  \quad & B(E2 $\uparrow$) (e$^2$ fm$^4$) & $\beta_2$\\
\hline
 QRPA  (SLy4) & 1.5 & 890 & 0.06 \\
\hline
 QRPA + REN &  0.9 &   2150 & 0.14 \\
\hline
 Exp.   & 1.2 & 2030 & 0.13 \\
\hline
\end{tabular}
\caption{Energy, reduced E2 transition strength and corresponding deformation parameter $\beta_2$ associated with the low-lying
 2$^+$ state of $^{120}$Sn, calculated according to QRPA and empirically renormalized QRPA as explained in the text, are
 compared to the experimental values \cite{Stelson}.
}
\end{table}

\begin{table}[h]
\begin{tabular}{|c|c|c|c|}
\hline
\multirow{2}{*}{Orbital}  & \multicolumn{2}{|c|}{$\varepsilon_\nu$ (MeV)} &  \multirow{2}{*}{$(m_{k})_{\nu}/m$}\\
\cline{2-3}

                       & SLy4 & Opt. & \\
\hline
 $d_{5/2}$   & -11.3 & -10.7 & 0.74 \\
\hline
 $g_{7/2}$  & -10.1 &  -10.5 & 0.67 \\
\hline
 $s_{1/2}$   & -9.0 & -7.9 & 0.80 \\
\hline
 $d_{3/2}$   & -8.5 &  -7.1 & 0.83 \\
\hline
 $h_{11/2}$  & -7.1 &  -7.45 & 0.67 \\
\hline
\multicolumn{3}{c|}{} & 0.74 ($\sigma = 0.07$) \\
\cline{4-4}
 
\end{tabular}
\caption{Energy of the valence orbitals associated with SLy4 ($m_{k} = 0.7m$) and those obtained by optimizing (Opt.) the NG quasiparticle spectrum to the data. In the last column the results labeled Opt. are parametrized in terms of an effective, state dependent 
$k$-mass.
}
\end{table}

\begin{table}
\begin{center}
\begin{tabular}{|c|c|c|c|c|c|c|}
\hline
& \multicolumn{2}{c|}{Exp.} & & & \multicolumn{2}{c|}{Th.} \\
\hline
$\epsilon_i$ (keV) & $\sigma$ (mb) & $d\sigma/d\Omega$  (mb/sr) && $\epsilon_i $  (keV) & $\sigma$ (mb)  & $d\sigma/d\Omega$  (mb/sr) \\
\hline
921&  0.63 &  0.75 && 1150 & 1.80 &  2.3 \\
1090& 5.35 &  7.0  &&1290 & 1.20 & 1.7 \\
1354& 1.66& 2.3 && 1710& 0.25 & 0.32 \\
1562 & 0.13 & 0.16  && 1910 & 2.90 & 4.0\\
1730 & 0.16 & 0.18  &&  & &\\
\hline
        &7.93 $\pm$ 2 & 10.39&  & & 6.15 & 8.32 \\
\hline\end{tabular}
\caption{\protect  The  most prominent experimental (theoretical) fragments of the $d_{5/2}$ single-particle state populated in the $^{120}$Sn(p,d) $^{119}$Sn $(5/2^+)$ reaction. 
The energies are listed in the first (fourth)  column, while the absolute cross sections  are given in the second (fifth), integrated  within 
the range $2^0 < \theta_{CM} < 55^0$,
and third (sixth), peak cross section, $(\theta_{CM})_{max} \approx 17^0$. The data are from \cite{Dickey}.} 
\label{fig:table2}
\end{center}
\end{table}

 \begin{table}
\begin{center}
\begin{tabular}{|c|c|c|c|}
\hline
  Observables  &  SLy4 &  $d_{5/2}$ shifted  & Opt. levels \\ 
\hline
$\Delta$                      &  10  (0.7\%) &  10  (0.7 \%) & 50   (3.5 \%)\\
 $E_{qp}$                     & 190 (19\%)   & 160  (16\%)   & 45  (4.5 \%) \\
 Mult.  splitt.               & 50  (7\%)    & 70  (10\%)    & 59  (8.4 \%) \\
  $d_{5/2}$ strength (centr.) & 200  (20\%)  & 40  (4\%)     & 40  (4\%) \\
$d_{5/2}$ strength (width)    & 160  (20\%)  &75  (9.3\%)    &  8  (1\%) \\
$B(E2)$                       & 1.4 (14\%)   & 1.34 (13\%)   & 1.43  (14\%) \\ 
$\sigma_{2n}(p,t)$            & 0.6 (3\%)    & 0.6 (3\%)     & 0.6 (3\%) \\
 \hline
\end{tabular}
\caption{\protect  Mean square deviation $\sigma$ between  the experimental data and the theoretical values taken at the  minimum of the corresponding functions displayed 
in Figs. 1(a) (inset),2(c,e,h),3(b) and 4(d) in keV for the pairing gap, quasiparticle energies, multiplet splitting, centroid and width of the  
$5/2^+$ low-lying single-particle strength distribution. In single-particle units $B_{sp}$ for the $\gamma$-decay  (B(E2) transition probabilities) and in mb for $\sigma_{2n}(p,t)$. In brackets
the ratio $\sigma/L$, called $\sigma_{rel}$  in the text,  between $\sigma$ and the experimental  range $L$ of the corresponding quantities: 1.4 MeV ($\Delta$), 1 MeV ($E_{qp}$), 700 keV (mult. splitting), 
1 MeV ($d_{5/2}$ centroid),  809 keV (=1730- 921) keV  ($d_{5/2}$ width), 10 $B_{sp}$ (B(E2)), 2250 mb ($\sigma_{2n}(p,t)$), is given.}
\label{fig:table4}
\end{center}
\end{table}

 \begin{figure}
		\includegraphics[width=0.9\textwidth]{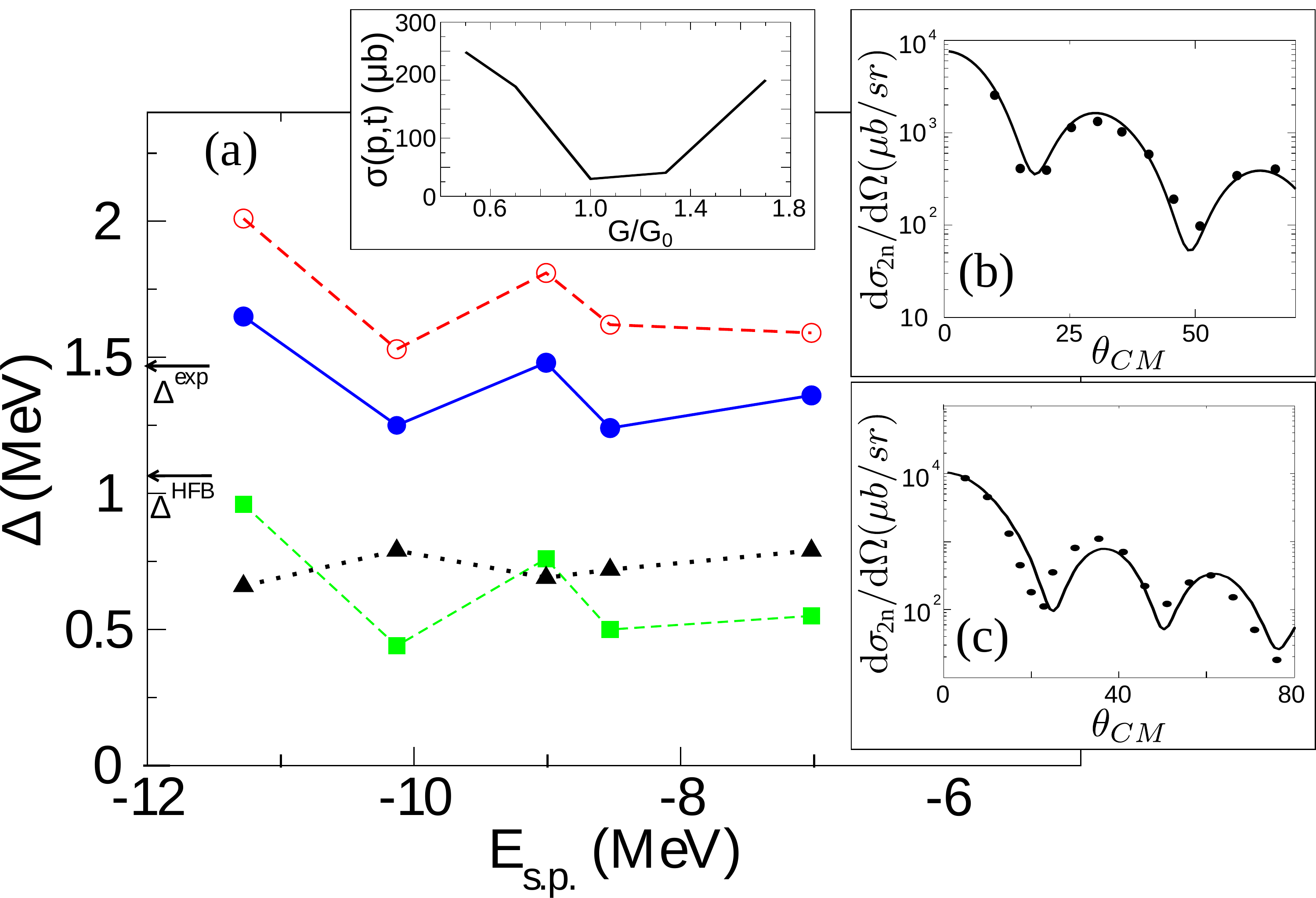}
	\caption{{\bf (a)} State-dependent pairing gaps for the five valence  orbitals of $^{120}$Sn (see Table II, second  column). 
The value of $\Delta$ associated with the HFB solution of $v_{14}(^1S_0))$ is indicated by an arrow labeled $\Delta^{HFB}$. 
 The pairing gaps calculated making use  of the empirically renormalised density modes 
 are shown  in terms of  open circles joined by a dashed line, 
 while the corresponding results obtained including  also spin modes, and thus corresponding to $\tilde \Delta_\nu$ are shown 
 by the solid dots joined by a continuous curve. 	
The  contributions $\tilde \Delta^{bare}_{\nu}$ and $\tilde \Delta^{ind}_{\nu}$ are displayed in terms of  solid triangles and solid squares joined by
dotted and by dashed  lines respectively. 
{\bf (b,c)} Calculated  two-particle transfer absolute differential cross sections associated with the reactions $^{120}$Sn(p,t)$^{118}$Sn (gs) 
	and $^{122}Sn(p,t)^{120}Sn(gs)$ (continuous curves)
	in comparison with experimental 
	data  (solid dots) \cite{Guazzoni1,Guazzoni2}.
	 In the inset of (a), the absolute value of the deviation of the integrated  theoretical absolute cross section from the experimental value in the case of the second reaction
	  is given as a function of the strength 
of the bare pairing interaction (cf. \cite{footnote2}).}
	\label{fig:2}
\end{figure}

\begin{figure*}
\includegraphics[height=0.18\textwidth]{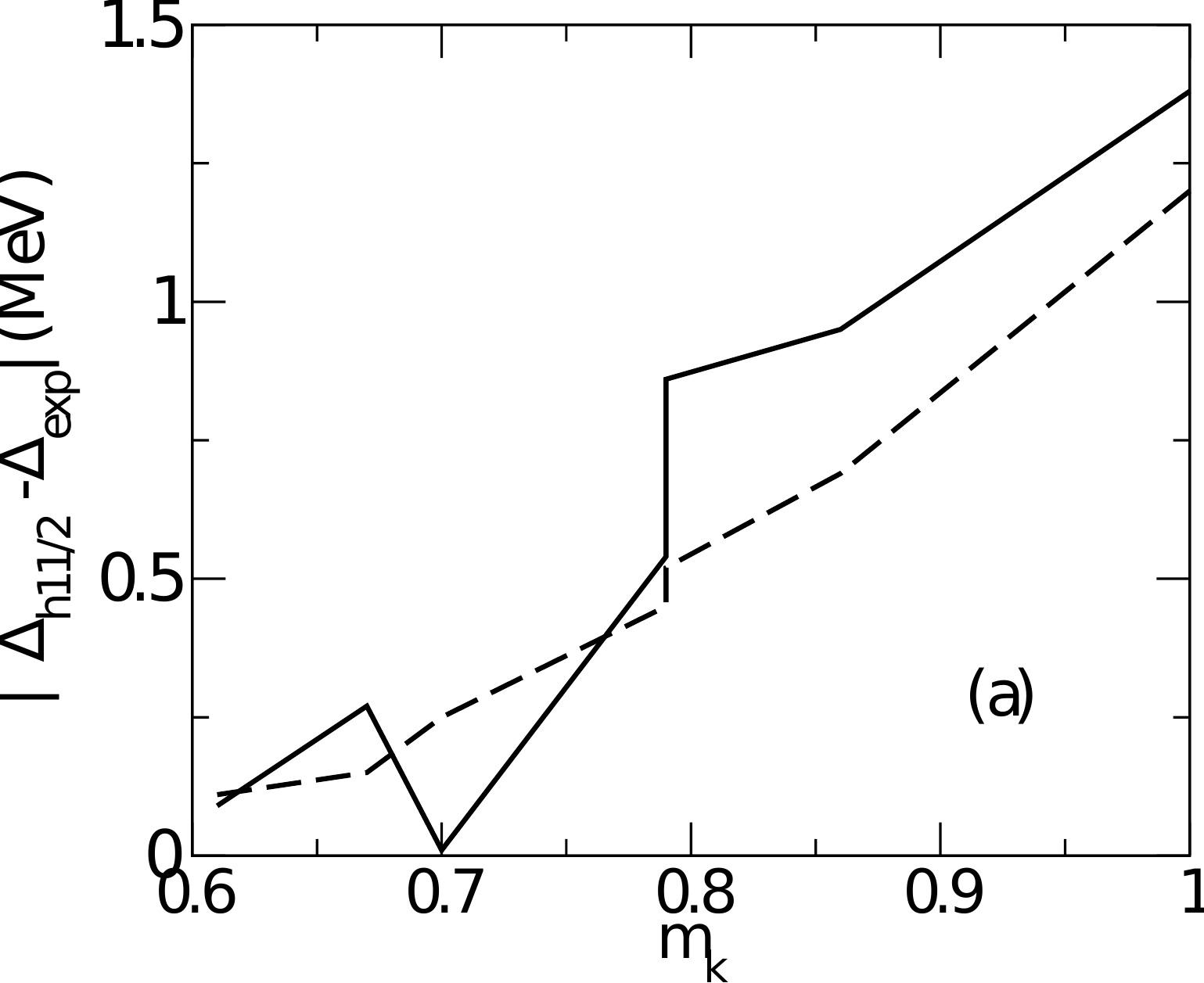}
\includegraphics[height=0.18\textwidth]{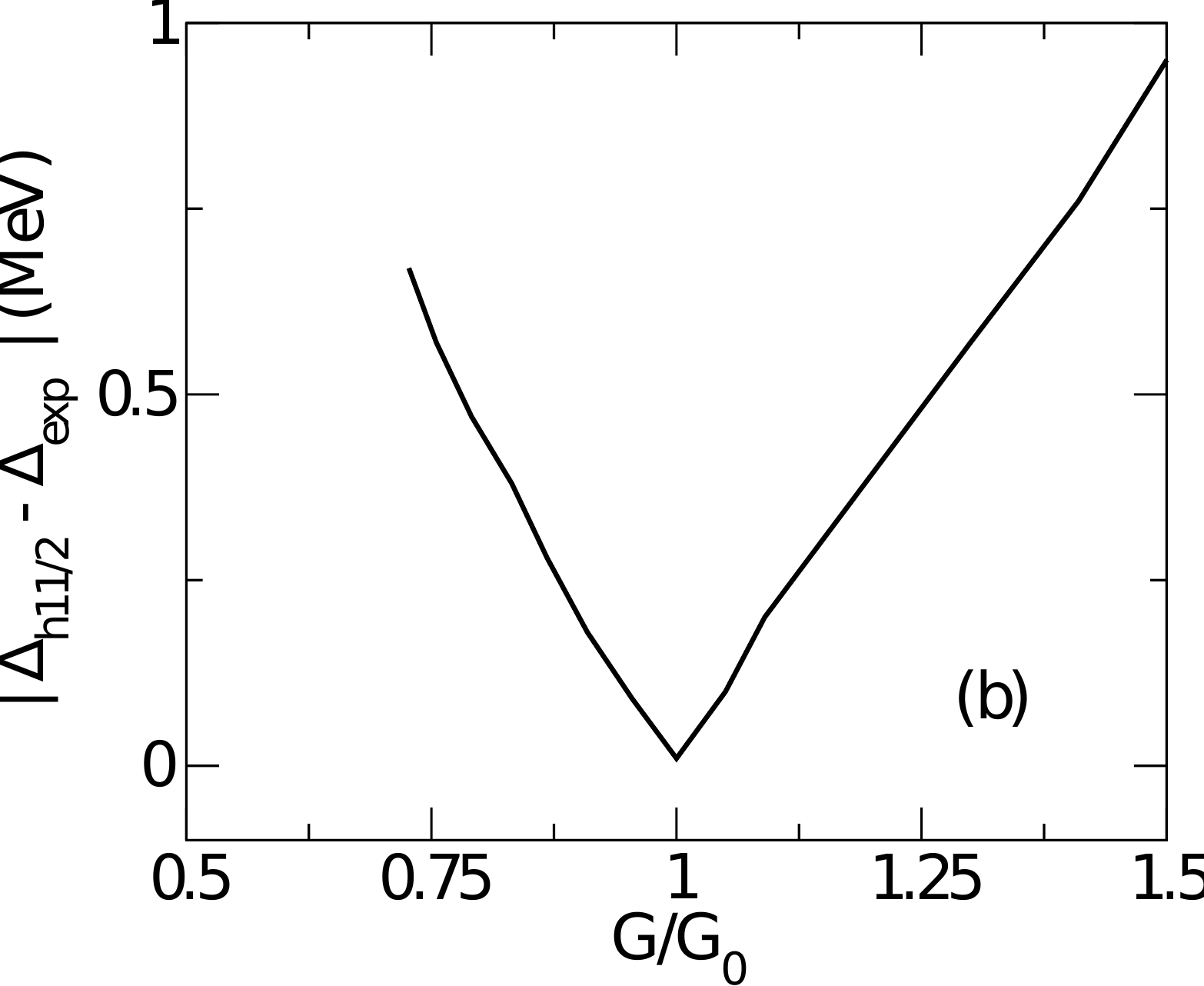}
\includegraphics[height=0.18\textwidth]{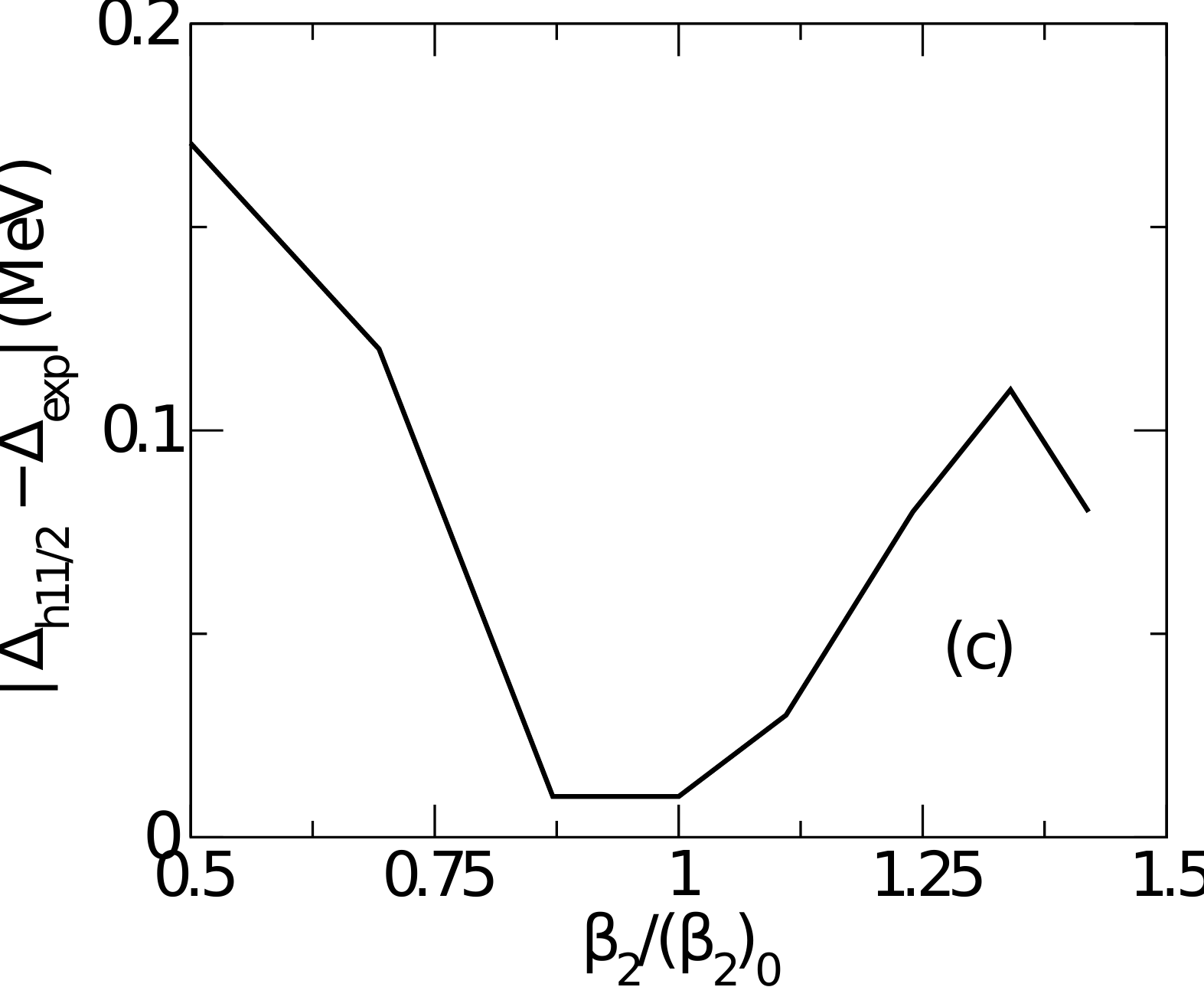}
\includegraphics[height=0.18\textwidth]{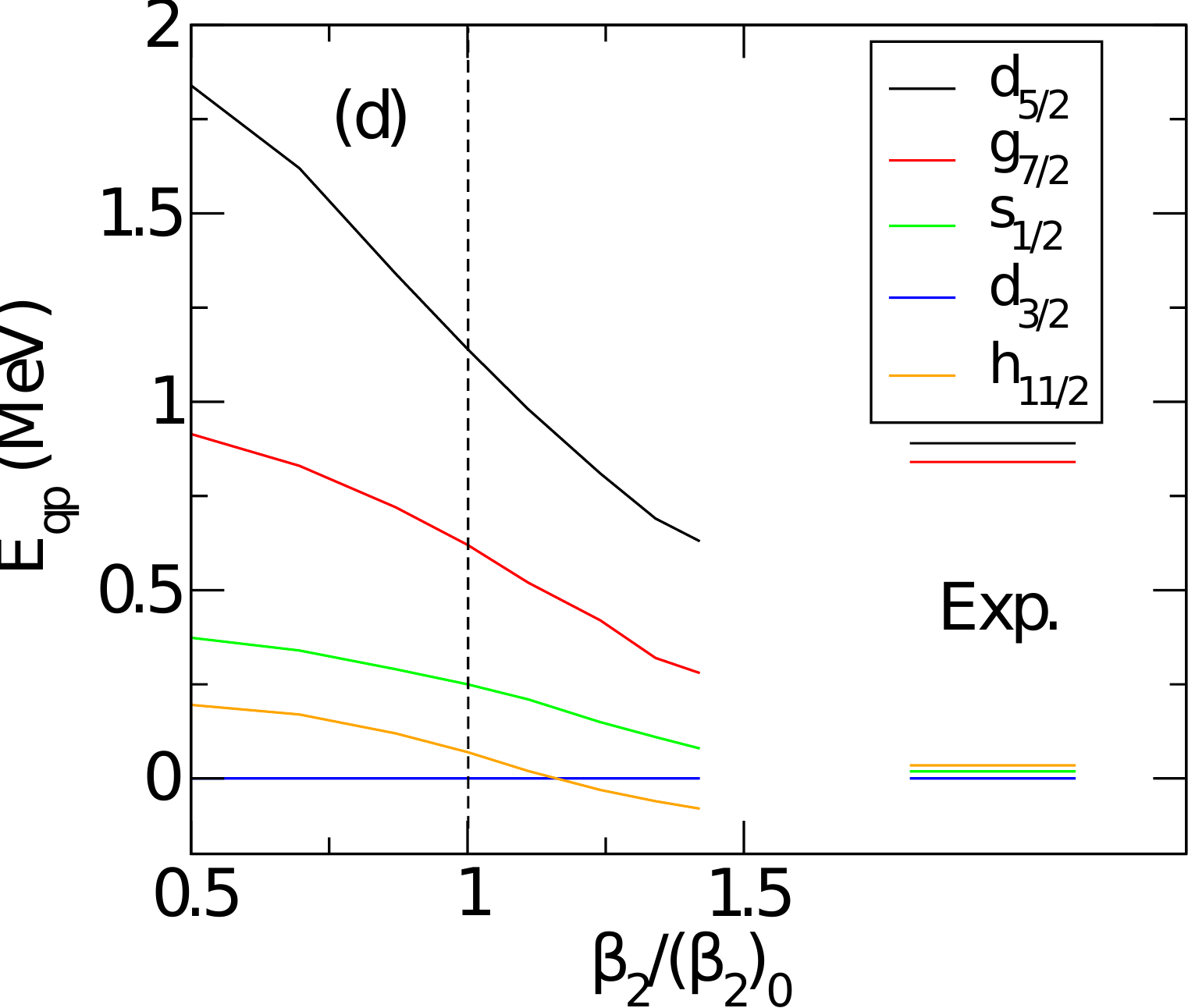}
\includegraphics[height=0.18\textwidth]{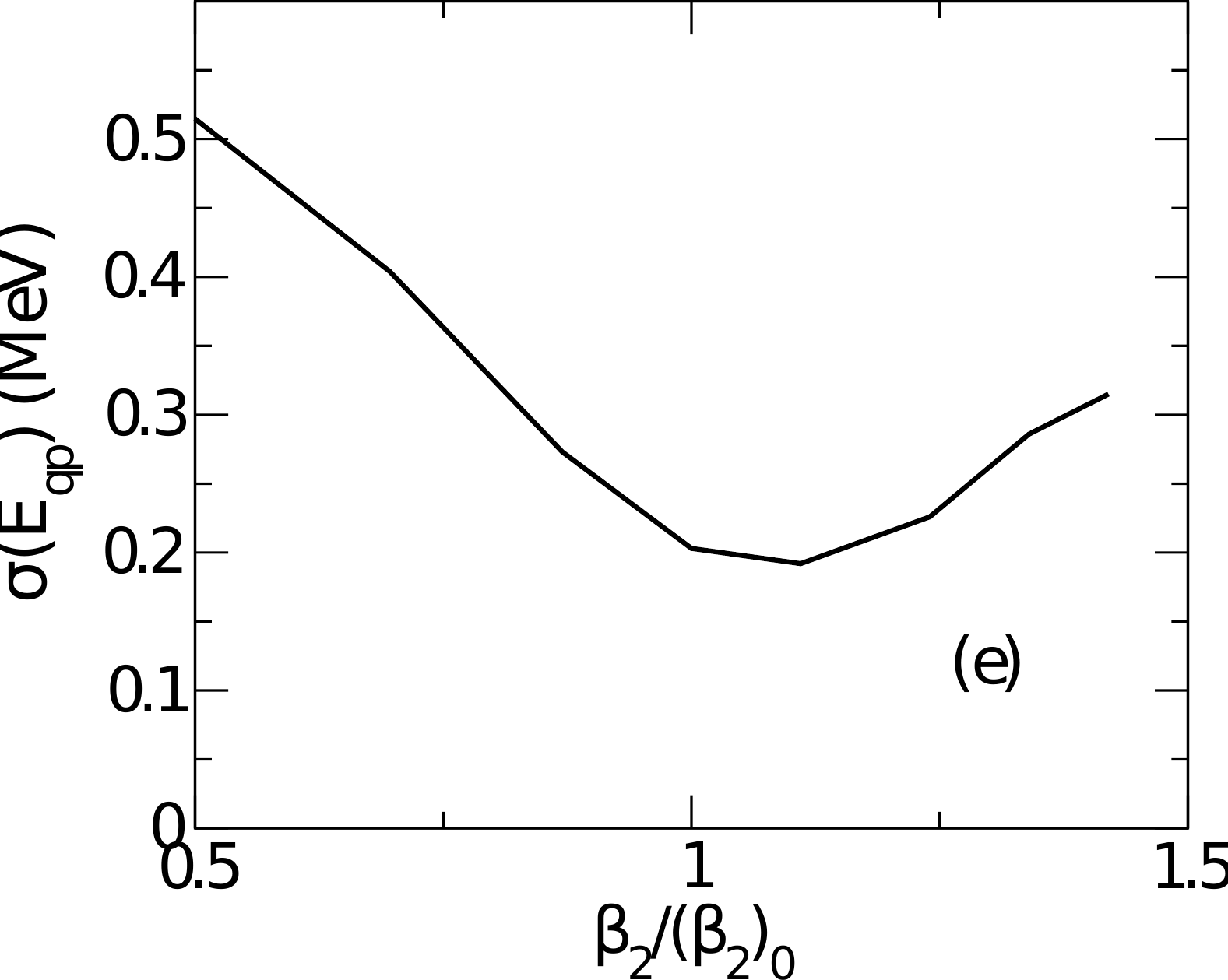}
\includegraphics[height=0.18\textwidth]{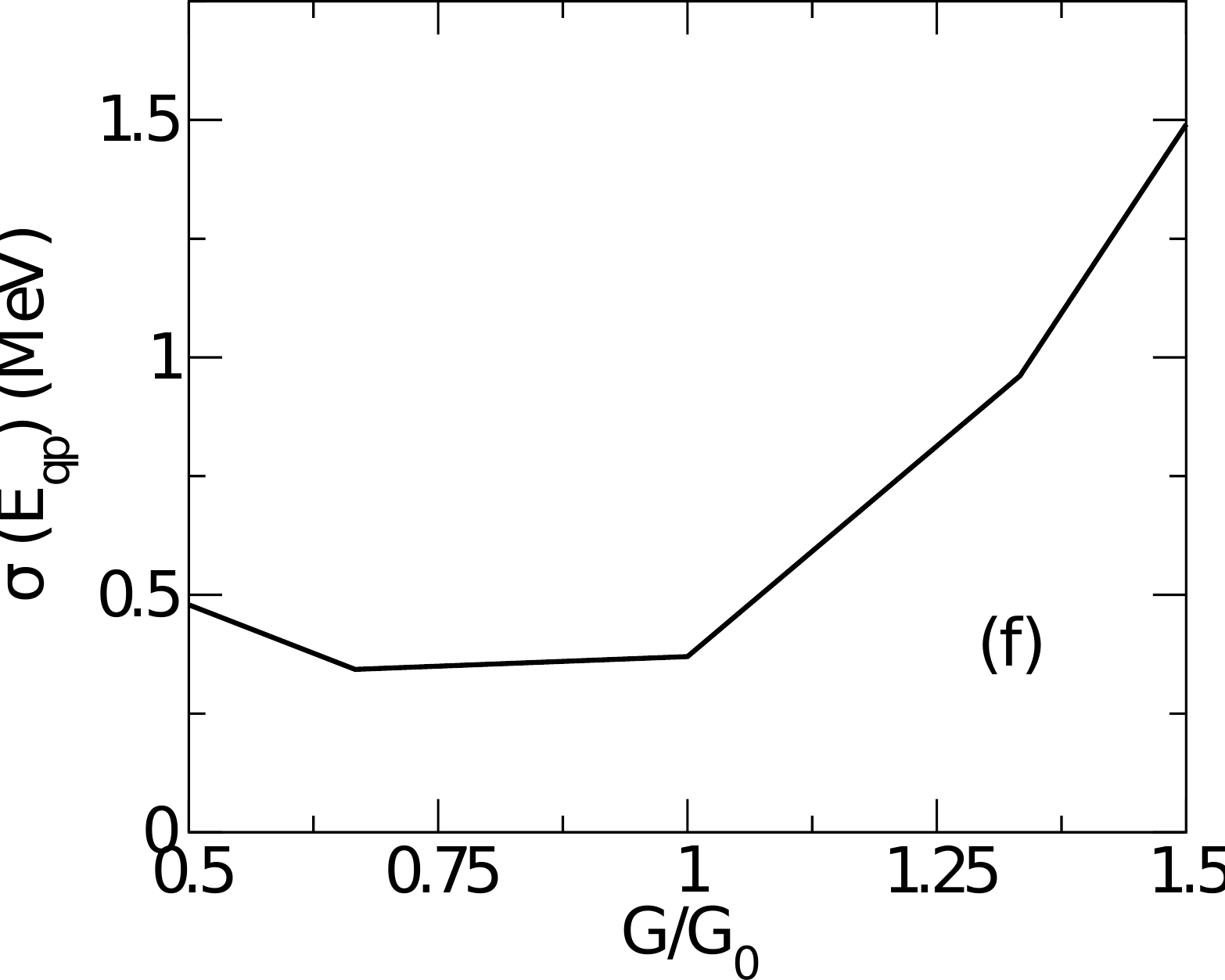}
\includegraphics[height=0.18\textwidth]{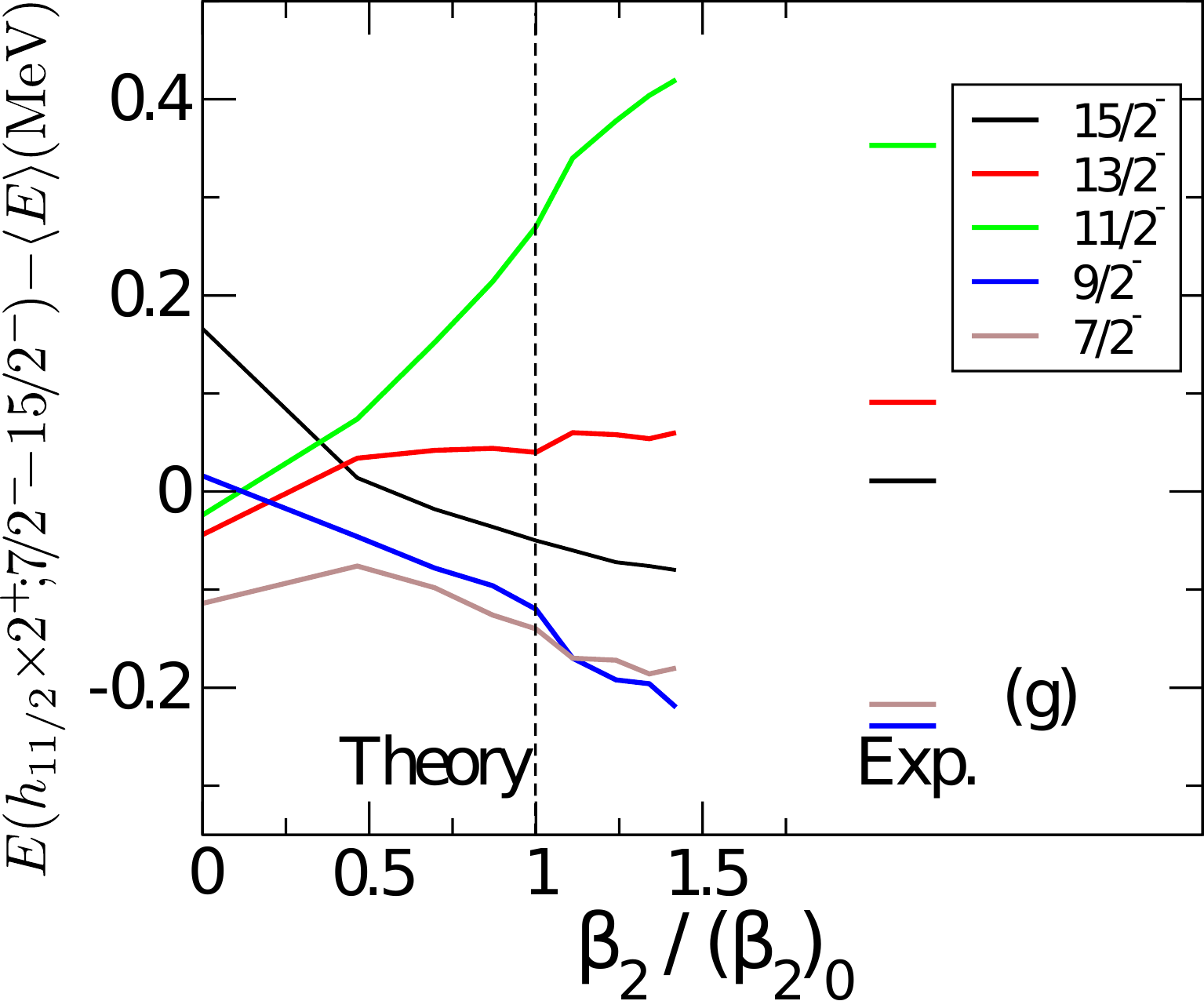}
\includegraphics[height=0.18\textwidth]{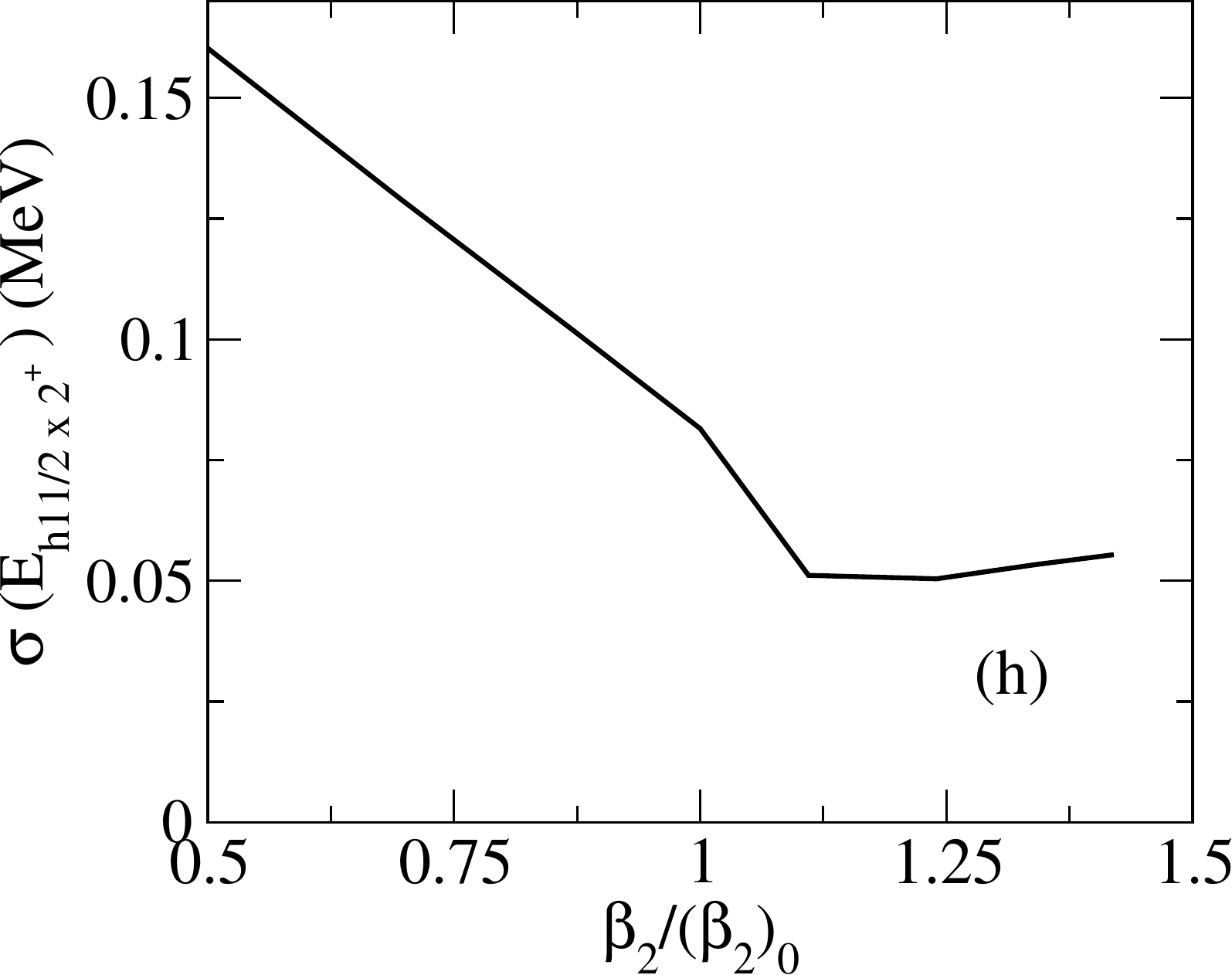}
\caption{\protect{
 Absolute value of the difference between the experimental pairing gap 
and the theoretical value of  $\Delta_{h_{11/2}}$ (see Eq.(1)),  calculated:
{\bf (a)} as a function of the effective mass $m_k$ associated with different Skyrme forces (solid curve) \cite{Skyrme}, the dashed curve displaying the gap 
obtained using  the fixed set of  valence levels $\epsilon_{\nu} (Opt.)$ (Table II);
{\bf (b)}  as a function of the ratio 
 $G/G_0$ 
  \cite{footnote2};
{\bf (c)} as a function of the ratio $\beta_2/(\beta_2)_0$ associated with the lowest quadruple mode of $^{120}$Sn  \cite{footnote_new}.
{\bf (d)} 
(color online) 
The lowest quasiparticle energy values  obtained from the full calculation as explained in the text 
referred to the energy of the $3/2^+$ state, in comparison with  the experimental data. 
{\bf (e)} Mean square deviation between the experimental and theoretical levels shown in (d).
{\bf (f)} Mean square deviation between the experimental and theoretical  energies of the five valence levels,
as a function of the ratio $G/G_0$. 
{\bf (g)} (color online)  The experimental  energies of the members of the   $h_{11/2} \otimes 2^+$ multiplet are compared with the 
theoretical values, calculated as a function of the ratio $\beta_2/(\beta_2)_0$.
{\bf (h)} Mean square deviation between  the experimental and theoretical  energies of the members of the  $h_{11/2} \otimes 2^+$ multiplet shown in (g).
 }}
\label{fig_single_particle}
\end{figure*}

\begin{figure}
\includegraphics[width=0.5\textwidth]{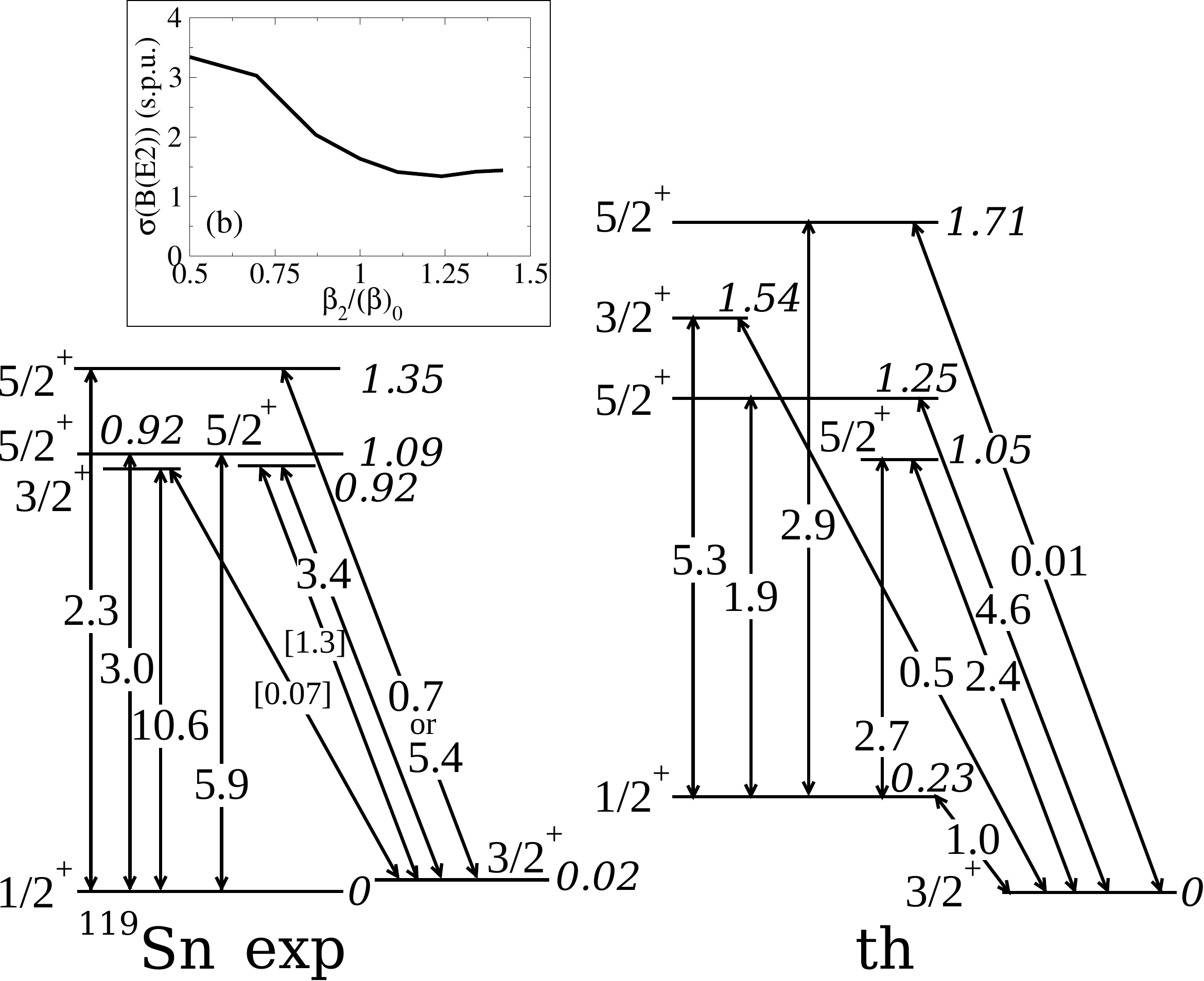}
\caption{ Experimental B(E2) values \cite{Stelson} in units of $B_{sp}$, of the quadrupole  $\gamma-$ decay following $^{119}Sn(\alpha,\alpha') ^{119}Sn^*$ Coulomb excitation,
 connecting the low-lying states  of $^{119}$Sn (left). 
 Also given are the theoretical values calculated 
making use of the results of the full renormalised calculation as explained in the text (right). The energies are in MeV.
Mean square  deviation between the  experimental transition strengths associated with E2 decay from the $5/2^+$ levels, and 
the theoretical values calculated as a function of the $\beta_2$ parameter  is given in the inset (upper left). }
\end{figure}

\begin{figure}
\includegraphics[width=0.9\textwidth]{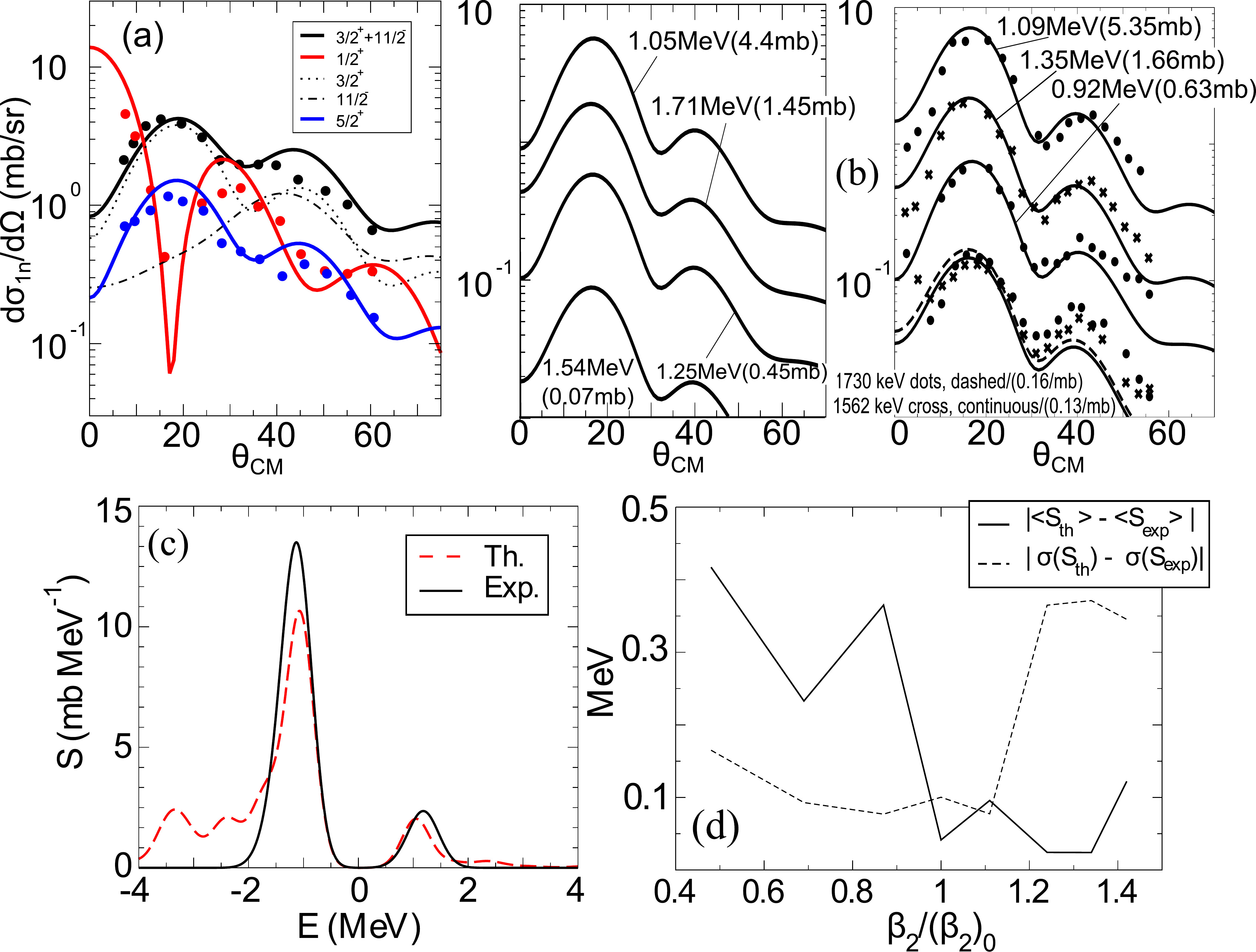}
\caption{ {\bf (a)} (color online) Absolute finite range, full recoil DWBA theoretical differential cross sections 
associated with the low--lying fragments of the $h_{11/2},d_{3/2},s_{1/2}$ and $d_{5/2}$ valence states most strongly populated in the 
reaction  $^{120}$Sn(d,p)$^{121}$Sn,  calculated with the help of state of the art optical potentials and $v_{np}$ interaction (I.J. Thompson, private
communication), making use of NG structure input,
in comparison with the experimental data \cite{Bechara}.
It is of notice that the $d_{5/2}$ single-particle orbit in the SLy4 mean field potential has been shifted towards  $\epsilon_F$ by 0.6 MeV (see text).
{\bf (b)} $^{120}$Sn(p,d)$^{119}$Sn $(5/2^+)$  absolute experimental differential cross sections \cite{Dickey}, 
together with the DWBA fit used in the analysis  of the data (right panel)
in comparison with the DWBA  calculations (left panel)  carried out as mentioned in (a) .
{\bf (c)} Comparison of the calculated strength function $S_{5/2} ( \sigma(^{120}Sn(p,d)^{119}Sn(5/2^+) + \sigma(^{120}Sn(d,p) ^{121}Sn(5/2^+))/E$
	with experimental data  derived from one-neutron transfer reactions \cite{Dickey,Bechara}. 
	The peaks have been folded together with a Gaussian function of variance  0.25 MeV. 
	{\bf (d)} The difference between the centroid  (width) of the experimental and of the calculated $d_{5/2}$ strength $S_{5/2}$ is shown as a function of the ratio 
	$\beta_2/(\beta_2)_0$ in terms of the solid  (dashed) curve.}
\end{figure}


\begin{thebibliography}{99}
\bibitem{footnote_new} This mode, used also in the calculation of the different quantities as a function of $\beta_2$,  is determined  as the QRPA solution 
of a separable multipole--multipole interaction with empirical single-particle levels, adjusting the strength 
to obtain the desired properties. One avoids in this way 
the difficulties associated with the zero--range character (ultraviolet divergencies) and finite size instabilities of most Skyrme forces
(cf. \cite{nft1,nft2} and refs. therein) . This procedure is what
is called empirical renormalization in the text.
\bibitem{Skyrme} In the calculation of the pairing gap $\Delta_{h_{11/2}}$ as a function of $m_k$,
the full  NG solution was worked out for the Skyrme interactions Ska ($m_k/m$= 0.61), SAMi (0.67), SLy4 (0.70),
SGII (0.79), SkM* (0.79), SkS1 (0.86) and SkP (1.0)  and the corresponding quantity  extracted
(continuous curve in Fig. 2(a)).
\bibitem{footnote2}  {In carrying out  the calculations reported in Figs. 2(b) and 2(f), namely the pairing gap $\Delta_{h11/2}$ and the deviation of the quasiparticle 
spectrum from experiment as a function of the strength of the bare $NN-$pairing force $v_{14}(^1S_0)$, 
the average value  of the matrix elements of this interaction  has been parametrised  in terms of a monopole-monopole
pairing force with constant matrix elements $G$. The reference value $G_0 (\approx 0.22 $ MeV)   has been  determined  so as  
to reproduce, in average, the $v_{14}(^1S_0)$  value of $\Delta_{j_{\nu}}$  
of the valence states. }
\end{thebibliography}
\end{document}